\documentclass[12pt]{article}

\usepackage{amsfonts,amssymb,amsbsy,latexsym,amsmath}

\usepackage[english]{babel}

\begin{document}
\large
\newcommand{\be}{\begin{equation}}
\newcommand{\ee}{\end{equation}}
\newcommand{\ba}{\begin{eqnarray}}
\newcommand{\ea}{\end{eqnarray}}
\newcommand{\dalam}{\raisebox{1mm}{\fbox{}{}}\;}
\newcommand{\pa}{\partial}
\newcommand{\s}{\sqrt}
\let\f\frac
\newcommand{\al}{\alpha}
\newcommand{\st}{\stackrel}
\newcommand{\tk}{\tilde\kappa}
\newcommand{\ep}{\epsilon}
\newcommand{\ds}{\displaystyle}
\newcommand{\ed}{\end{document}}
\newcommand{\ul}{\underline}

\begin{center}
{\bf Explicit Chiral Symmetry Breaking as a Premise of the Cross-Sections' Rise}\\

\vspace*{0.2cm}

 V. A. Petrov\\

\vspace*{0.2cm}

{\it Division of Theoretical Physics,
Institute for High Energy Physics,
142 281 Protvino, RF}

\end{center}

\vspace*{0.4cm}

\begin{abstract}

We argue that if QCD yields a theory of interacting hadrons then
explicit chiral symmetry breaking is a necessary condition for
infinitely rising cross-sections. Otherwise cross-sections go to
zero at high energies.

\end{abstract}

    On the eve of the LHC operation a special consideration  is given to 
   measurements of the total and elastic
   cross-sections in $pp$ collision at yet unattainable energies up 
   to $10-14\,TeV$ in the c.m.s.
   All previous measurements since pioneering studies of $K^{+}p$
   interactions at $E_{lab} = 30-60\,GeV$ [1] have shown a steady rise both of
   total and elastic cross-sections till the Tevatron and RHIC energy region.
   Cosmic rays data (large errors though) also witness in
   favor of rising cross-sections at least up to several tens TeV.
   One could ask a somewhat academic question: "If such a rise will
   go on infinitely?" Certainly, it seems almost evident that at
   high energy enough when it will not make sense to distinguish
   among different forces (and now we prescribe the rise in question
   to the strong interaction) the situation can drastically change.
   Nonetheless it seems to be of some use to try to understand
   general features of QCD as the underlying theory for strong
   interactions of hadrons.
    In the low-energy sector it proves useful to consider even
    quarkless theory as a training ground for the study of confinement,
    estimates of the glueball masses etc (mostly in lattice QCD).
    Another approximation, QCD with massless
    quarks, also proves to be very close to reality in the sector of light
    hadrons. For instance, the nucleon mass changes insignificantly
    when adding current quark mass terms [2].
    So, it is tempting to assume that at high energies these
    approximations will work at least not worse.

    We will try to check such an assumption [3]. If we deal with
    Quantum Gluodynamics which supposedly results in a theory of
    colorless massive hadrons~ -~ glueballs ~- ~interacting with each
    other, then we are with a single mass scale  $\Lambda_{\mbox{\small {QCD}}}$,
    a clear imprint of the "dimensional transmutation". All glueball
    masses (including resonances) are multiples of the fundamental scale

    $$
M_i^2 = c_i \Lambda^{2}_{\mbox{\small {QCD}}},
$$
    Coefficients $c_i$ are pure numbers. To reveal the provenance of the
    fundamental scale $\Lambda_{\mbox{\small {QCD}}}$   from the 
    underlying theory we recall that
    from the renormalization group considerations we have the
    following relationship with coupling $\alpha_s$ and mass scale $\mu$ (renormalization scale)
    
$$
 \Lambda^2_{\mbox{\small {QCD}}}=  \mu^2\exp(-K(\alpha_s)),  \qquad \qquad  \qquad (1)
$$

$$
dK(\alpha_s)/d\alpha_s = 1/\beta(\alpha_s),
$$

$$
K(\alpha_s) \sim 1/\beta_0\alpha_s + O (\log (1/\alpha_s))\;\;
\mbox{at}\;\; \alpha_s \to 0.
$$
    Nonanalytic dependence at small  $\alpha_s$ is
    similar to the superconductivity energy gap coupling
    dependence.
    In the present context one may argue that
     parameter $\Lambda_{\mbox{\small {QCD}}}$  is
    scheme dependent. However, from the evident property of the
    physical hadronic amplitudes to be both RG and scheme invariant we
    are free to use any convenient scheme. In the same way we use
    a convenient frame for relativistic invariant quantities.
    We can also refer to paper  [4]   where a scheme independent expression
    for the parameter $\Lambda_{\mbox{\small {QCD}}}$ has been obtained.
     Let us  look now what happens if we take the free-field
     limit $\alpha_s \to 0$. As is easily seen from Eq.(1), all masses go to
     zero. Physically it means that "glueballs" degenerate into a
     set of systems of collinear  gluons which are evidently massless.
     On the other hand it is also evident that any
     scattering amplitude vanishes in the free-field limit: no
     interaction occurs.
     Take for example an elastic scattering amplitude at $t = 0$.
     We remind that in the confined gluodynamics no massless hadrons
     take place (due to the trace anomaly), so that  $t = 0$   is an analyticity point.
     From the dimensional reasons the amplitude is of the form

$$
T (s, 0) = \Phi (s/\Lambda^2_{\mbox{\small {QCD}}}; \{c_i\}),
$$

     \noindent where $\{c_i\}$ are pure numbers , in general complex, and encode the
     physical spectrum and depend on spins etc.
     The rest of the argument is almost trivial. One can see that
     the free-field limit is identical to the infinite energy
     limit.
     Therefore the amplitude vanishes at infinite energies.
     If we take fixed-angle high energy scattering then it is easy
     to see that we come to the same conclusion. At fixed and non-zero $t$
     (in the physical region) one can see that for the imaginary part of the amplitude the
     conclusion about decreasing holds because

$$
Im T (s, t)\leq Im T (s, 0).
$$
As to the real part of the amplitude we can only refer to
paper [4] where it has been proved that (at least for some sequence\\
of $s$)
$$
 |T (s, t)|\leq |T (s, 0)|
$$
\emph{if} the even signature amplitude dominates forward scattering.
However the authors of [4] assumed that the total cross-sections can
decrease not faster than $s^{-1/2}$ while, in our case, they decrease
faster than $s^{-1}$.

      One can see that inclusion of massless quarks does not
     change the conclusion about the decrease of the amplitudes.
     The only difference is that , due to spontaneous chiral symmetry
     breaking, massless Goldstone bosons appear (pions etc)which
     can in principle spoil analyticity at $t = 0$. One can show,
     however, that, say, the imaginary part of the pi-nucleon scattering amplitude
     remains finite at $t = 0$. So, one can consider pi-nucleon total cross-sections
     which  asymptotically drop to zero as well.
     We have to  note also that non-zero chiral condensate $\langle\bar{\psi}\psi\rangle$
     (which can dynamically generate quark masses)does not introduce a new scale, independent of
$\Lambda_{\mbox{\small
     {QCD}}}$ and cannot invalidate the argument.

      The situation changes  radically if chiral symmetry is
     broken explicitly i.e. the Lagrangian contains terms
$$
m \bar{\psi}\psi
$$

Now we have got two independent RG-invariant mass-scales, the old
one,
 $\Lambda_{\mbox{\small
     {QCD}}}$  and a new one which can be chosen as

$$
 M^2 = m^2\exp(L(\alpha_s))
 $$

\noindent where

$$
dL(\alpha_s)/d\alpha_s = \gamma_m (\alpha_s)/\beta(\alpha_s)
$$

\noindent and $\gamma_m (\alpha_s) =- d(\ln m^2)/d\ln \mu^2$ is the "mass
anomalous dimension" defined as in [6]. For simplicity we take one
flavor. With two mass scales the forward scattering amplitude is of
general form

$$
T(s, 0) = F (s/ \Lambda^2_{\mbox{\small
     {QCD}}}, M^2/ \Lambda^2_{\mbox{\small
     {QCD}}}).
$$
At infinite energy $T \to  F (\infty, M^2/ \Lambda^2_{\mbox{\small
     {QCD}}})$,
while at $\alpha_s \to 0$,   $T \to  F (\infty, \infty) = 0$ because
$M^2 \sim (\frac{1}{\alpha_s})^{\frac{\gamma_{m_0}}{\beta_0}}$
at $\alpha_s \to  0$. So the (massive) free-field limit is generally
different from the high-energy limit and we cannot come to any
definite conclusion concerning the latter. Nonetheless, one thing is
clear: the infinite rise of cross sections with energy is
impossible, in the framework of QCD, without an explicit chiral
symmetry violation, i.e. without  current quark mass terms in the
Lagrangian.

 Normally the origin of these current quark mass terms is associated
with electroweak symmetry breaking and Yukawa couplings of quarks
to the Higgs field. From this angle, the rise of hadronic total
cross-sections seems to be strangely induced by the electroweak part
of the SM. In other words, the rise of hadronic cross-sections is
not a purely strong-interaction feature.

\vspace*{0.2cm}

{\bf Acknowledgements.} 

I am very much indebted to Gia Djaparidze, Anton Godizov,
Alexandre Kisselev, Andr\'e Martin, George Pron'ko, Vladimir Rochev, Sergey Troshin and
Tai Wu for useful discussions of this subject.

\end{document}